\documentclass[12pt]{book}
\usepackage{plenum}
\usepackage{psfig}
\begin{document}


\chapter{THE E895 $\pi^-$ CORRELATION ANALYSIS -- A STATUS REPORT}

\vspace*{-1cm}

\author{
M.A.~Lisa\refnote{j}, for the E895 Collaboration\footnote{Presented at the $14^{\rm th}$
Winter Workshop on Nuclear Dynamics, Snowbird, UT, Feb 1998}
\newline
N.N.~Ajitanand\refnote{m},
J.~Alexander\refnote{m},
D.~Best\refnote{a},
P.~Brady\refnote{e},
T.~Case\refnote{a},
B.~Caskey\refnote{e},
D.~Cebra\refnote{e},
J.~Chance\refnote{e},
I.~Chemakin\refnote{d},
P.~Chung\refnote{m},
V.~Cianciolo\refnote{i},
B.~Cole\refnote{d},
K.~Crowe\refnote{a},
A.~Das\refnote{j},
J.~Draper\refnote{e},
S.~Gushue\refnote{b},
M.~Gilkes\refnote{l},
M.~Heffner\refnote{e},
H.~Hiejima\refnote{d},
A.~Hirsch\refnote{l},
E.~Hjort\refnote{l},
L.~Huo\refnote{g},
M.~Justice\refnote{h},
M.~Kaplan\refnote{c},
D.~Keane\refnote{h},
J.~Kintner\refnote{f},
D.~Krofcheck\refnote{k},
R.~Lacey\refnote{m},
J.~Lauret\refnote{m},
E.~LeBras\refnote{m},
H.~Liu\refnote{h},
Y.~Liu\refnote{g},
R.~McGrath\refnote{m},
Z.~Milosevich\refnote{c},
D.~Olson\refnote{a},
S.~Panitkin\refnote{h},
C.~Pinkenburg\refnote{m},
N.~Porile\refnote{l},
G.~Rai\refnote{a},
H.-G.~Ritter\refnote{a},
J.~Romero\refnote{e},
R.~Scharenburg\refnote{l},
L.~Schroeder\refnote{a},
R.~Soltz\refnote{i},
B.~Srivastava\refnote{l},
N.T.B.~Stone\refnote{b},
T.J.~Symons\refnote{a},
S.~Wang\refnote{h},
R.~Wells\refnote{j},
J.~Whitfield\refnote{c},
T.~Wienold\refnote{a},
R.~Witt\refnote{h},
L.~Wood\refnote{e},
X.~Yang\refnote{d},
W.~Zhang\refnote{g},
Y.~Zhang\refnote{d}
\\
\vspace{2mm}
\refnote{a}LBL, \refnote{b}BNL, \refnote{c}CMU, \refnote{d}Columbia,
\refnote{e}UC~Davis, \refnote{f}St.~Mary's~College, \refnote{g}Harbin~Institute,~China,
\refnote{h}Kent~State, \refnote{i}LLNL, \refnote{j}Ohio~State, \refnote{k}Auckland,~NZ,
\refnote{l}Purdue, \refnote{m}SUNY~at~Stony~Brook
}

\section{INTRODUCTION-- E895 MOTIVATION AND EXPERIMENT}

A primary goal of high-enery heavy ion physics is to create and
study the quark-gluon plasma (QGP), a phase of matter
in which partonic-- instead of hadronic-- degrees of freedom describe the system.
Several transport, hydrodynamic, and nucleation
theories\refnote{\cite{BALi,Kapusta,earlyRischke,glen}} suggest that energy
densities achieved in central collisions between heavy ions
at AGS energies may be sufficient to create the QGP.

If the QGP is created in a heavy ion collision, the timescale for particle
emission is expected to be longer than a scenario in which only ordinary
hadronic degrees of freedom play a role\refnote{\cite{pratttimescale,RischkeHBT}},
due to the extra time of hadronization.
Thus, one proposed signature for QGP formation has been a large apparent 
source lifetime as measured by pion HBT measurements.
However, HBT analyses of very heavy ion collisions
at the maximum AGS energy\refnote{\cite{E877HBT,E802HBT}}
(10.6~AGeV) and at CERN SPS\refnote{\cite{NA44,NA49HBT}} (158~AGeV)
do not indicate emission timescales longer than that expected from normal
hadronic physics.  Thus, it may seem pointless to look for long lifetimes
at energies {\it below} maximum AGS energy.

However, recent hydrodynamical calculations by Rischke and collaborators
suggest that some signatures of QGP creation-- including large source
lifetimes from HBT--
may {\it only} be apparent very close to the threshold of QGP formation
\refnote{\cite{RischkeHBT}}.  The QGP threshold energy
corresponds to a ``softest point'' in the Equation of
State\refnote{\cite{RischkeHBT,Shuryak,Rischke1}}.  For
a source created at this energy, the lifetime is longer because
the system does not expand and cool as rapidly as it would if there
were no phase transition.  For collisions at energies much above this
threshold energy, the system (which is in the QGP phase) expands and cools
rapidly, and the lifetime effect is diminished.

The value of the threshold energy, then, is of paramount importance.
One would like to study collisions around this energy, to better 
recognize and understand the transition.  Since this value is not known
in principle, it is important to perform systematic studies of nuclear
collisions as a function of energy.

Further motivation to look for QGP turn-on at the AGS comes from 
recent thermochemical meta-analyses\refnote{\cite{PBMJS}} of experimental
spectra and yields from collisions at maximum AGS and SPS energies.
Based on an equlibrium scenario, these analyses suggest that already
at the maximum AGS energy, the system freezes out 
on the border between QGP and normal nuclear matter.  This would imply that
the system had cooled from a hotter, denser state in the QGP phase.  Similar
analyses in the SIS/Bevalac energy region\refnote{\cite{FOPIRomania}}
(0.1-1.0~AGeV) place the systems created at these energies solidly in the
realm of normal hadronic matter, but smoothly approaching the
hadronic-matter/QGP ``border'' (in the phase diagram) as
the bombarding energy increases.



Using the Time Projection
Chamber\refnote{\cite{TPCLBLreport,EOSieee}} used in the EOS experiments
at the Bevalac,
the E895 collaboration has measured roughly 0.5-1 million collision events
at 2, 4, 6, and 8~AGeV at the Brookhaven AGS.  The results presented
here represent a small ($\sim$2\%)
fraction of the total available statistics.

The TPC was located in the
MPS magnet operated with a field of 0.75 or 1 T.  The active volume of the
TPC is a rectangular region 154~cm~x~96~cm~x~75~cm in the beam, bend, and drift
directions, respectively.
Electrons liberated by charged particles passing through the
TPC drift to 15360 pads arranged in 128 rows at the bottom of the TPC.
The signal on each pad is sampled and digitized every 100 ns (140 time buckets),
providing roughly 2 million 3-dimensional pixels in which the ionization
is measured.  ``Hits'' are reconstructed from the pixels, corresponding
to a track crossing a padrow.  Tracks are then formed from the found hits,
giving
continuous tracking and particle identification with nearly 4$\pi$ acceptance
in the center of mass.

\section{HBT ANALYSIS AND THE NEED FOR PAIR-WISE CUTS}

\newcommand{\kvec}[1]{\mbox{$\vec{k}_{#1}$}}

The correlation function C(\kvec{1},\kvec{2})
is given by constructing the ratio

\begin{equation}
C(\kvec{1},\kvec{2}) = \frac{R(\kvec{1},\kvec{2})}{B(\kvec{1},\kvec{2})}
\label{eq:correl}
\end{equation}

\noindent
where \kvec{1}\ and \kvec{2}\ are the momenta of the two
particles (here, pions) in a pair. $R$ 
is the measured (``real'') 2-particle yield.  The background yield 
$B$ should contain all phase space and
single-particle detector acceptance effects. It is constructed via the
event-mixing technique; we mix $\pi^-$ from a given event with pions
from the previous 15 events.

In one-dimensional HBT analyses, such as the one discussed here, the
real and background distributions are binned
in $Q_{inv}$, where $Q_{inv}^2 = (\kvec{1}-\kvec{2})^2 - (E_1-E_2)^2$.
 $C(Q_{inv})$ is normalized to unity at large $Q_{inv}$.  All correlation
functions presented here are binned in $Q_{inv}$ in GeV/c.

Three distinct ``levels'' of cuts are applied to the data input to the HBT analysis.
Firstly, event-wise cuts are applied, to select a range of charged
particle multiplicity, and a range of primary vertex positions (the latter
helps eliminate events from upstream of the target).  In the current analysis,
the multiplicity range for the 2 and 4~AGeV events selected was large
(50-200 and 50-250, respectively, corresponding to a maximum impact parameter
of about 8 fm), while the 8~AGeV data was more central (multiplicity$\approx$200-300,
or b$\approx$0-3 fm).

Secondly, track-wise
cuts are performed.  These include selection of particle type, goodness of
the track fit, track length, and phase space (\kvec{}) cuts.  In most
E895 analyses, the most crucial track quality cut has been on how well
the track projects back to the primary vertex.  These cuts are applied to
events and particles both in the measured yield $R$ as well as the
background $B$.

Below, we discuss the need for a third level of cuts-- pair-wise cuts--
to reduce two-particle acceptance effects.

\subsection{Split tracks}

\begin{figure}[tbp]
\centerline{\hbox{
\psfig{figure=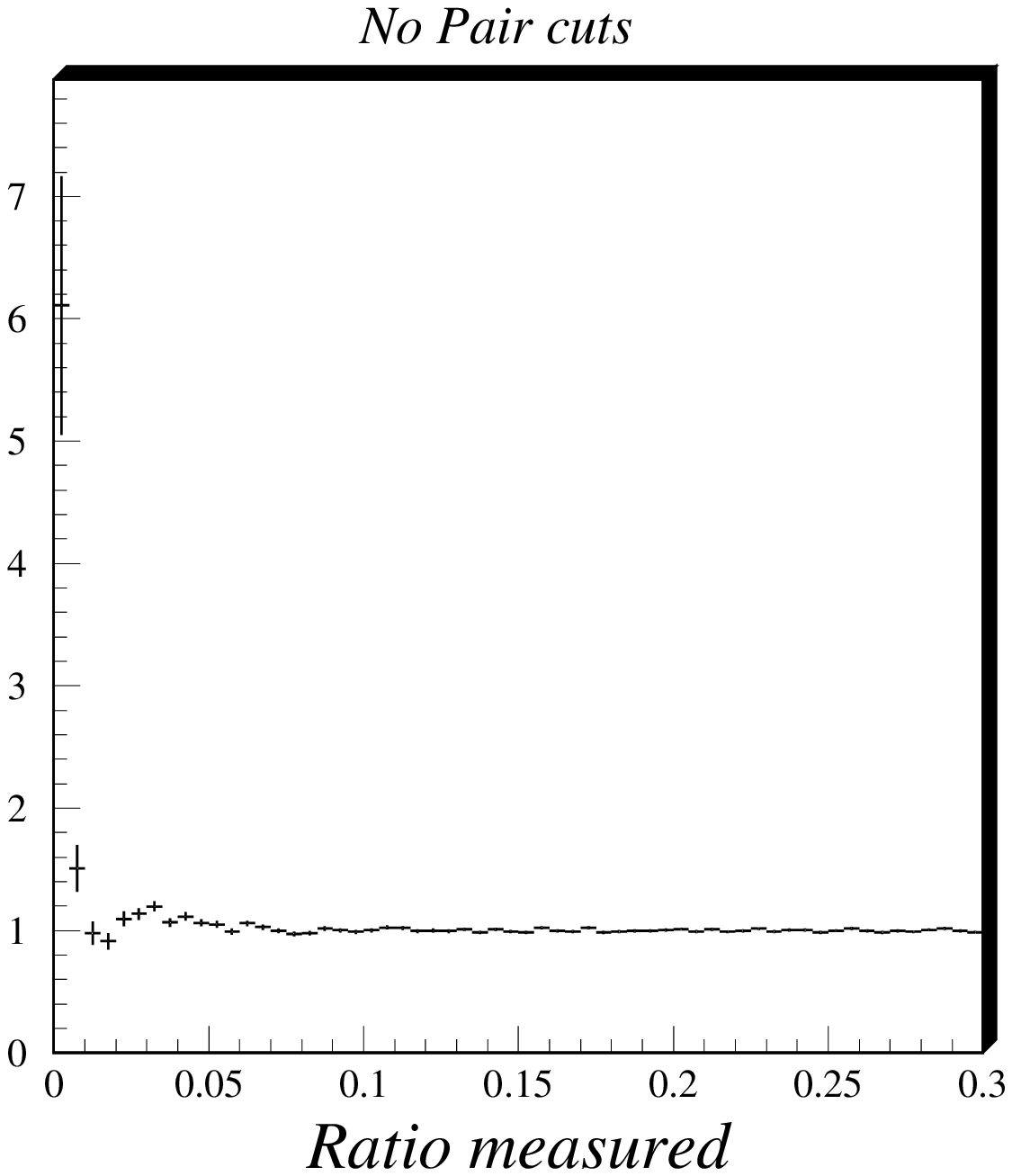,height=2.6in}
\psfig{figure=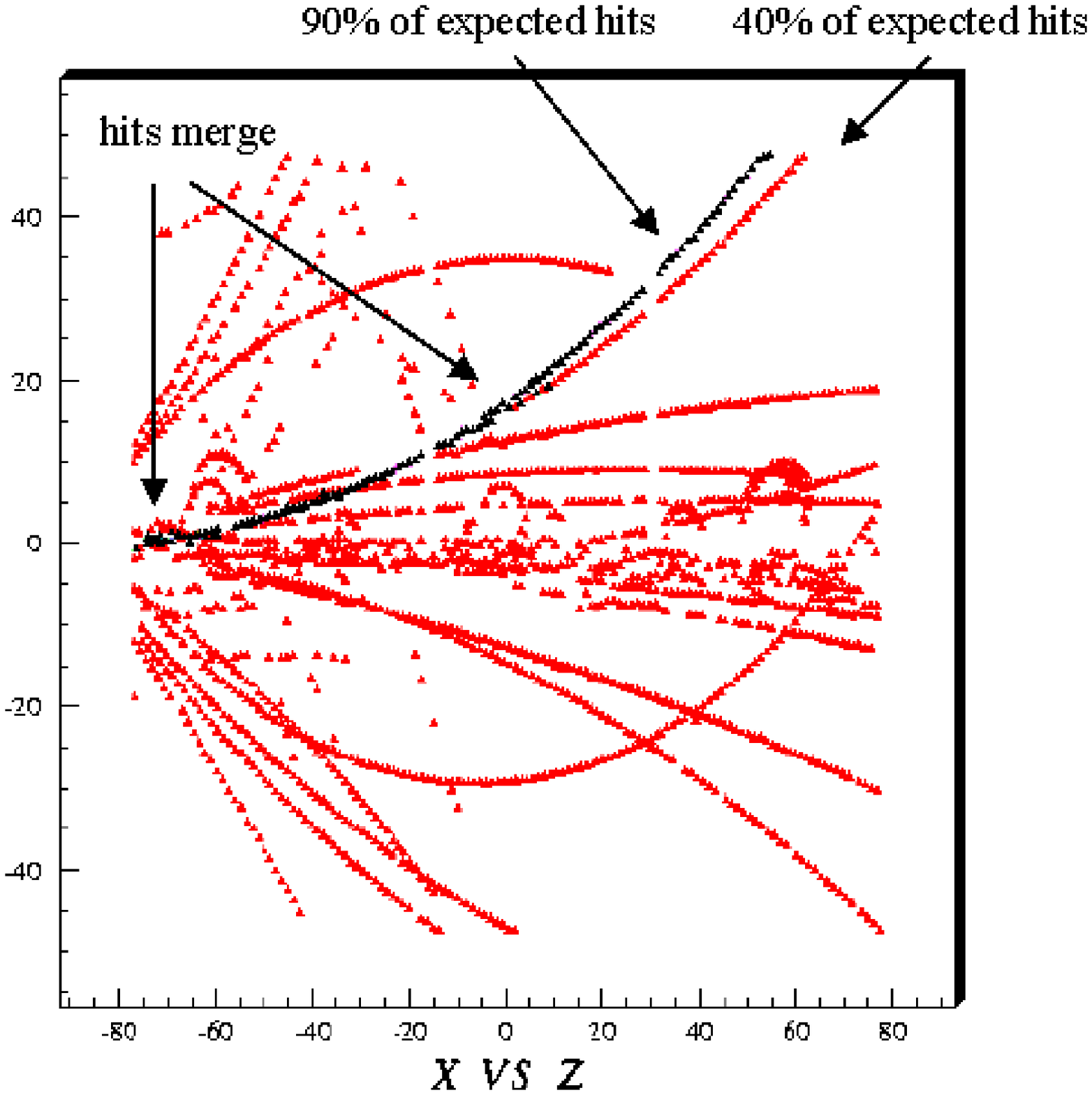,height=2.6in}
}}
\caption{{\bf Left} The correlation function with only event-wise and track-wise
cuts applied.  Structure at low $Q_{inv}$ is due primarily to track-splitting and merging.
\newline
{\bf Right} Hits for true low-$Q_{inv}$ pair in a very low multiplicity event.
Hits close to the primary vertex merge, and are assigned to one of the tracks.}
\label{fig:nopaircuts}
\end{figure}

Using only track-wise cuts to select ``good'' pions, the measured correlation
function for the 4~AGeV data is shown in the left panel of Figure~\ref{fig:nopaircuts}.
Simulations and visual inspection of individual events indicate that
the strong and unrealistic structure seen at low $Q_{inv}$ is a result of
track splitting.  Here, a track which crosses, say, 80 padrows (and so in
principle should produce 80 ``hits'' (see above)), is broken by the pattern
recognition software into two tracks with 20 and 35 hits. (Overall loss of 30\%
of the hits on a track-- even those not split-- is typical, due to the
high track density and hit merging.)  Naturally, the
reconstructed momentum difference for this false pair is low.  Such pairs will
be seen in the ``real'' distribution $R$, but not in the background $B$.
Track splitting mainly affects the lowest $Q_{inv}$ bin, but the effect extends
to $\sim$30 MeV/c.

A possible track-wise cut that can remove this effect would be to require that
more than 50\%
of a track be reconstructed.  The problem with this approach is seen in the
right panel of Figure~\ref{fig:nopaircuts}.
A large fraction of ``true'' low-$Q_{inv}$ pairs are eliminated as well, since, close to
the primary vertex, the tracks are closer than the two-hit resolving distance
($\sim$1.5 cm), so only one hit is found;
this hit is assigned to one of the tracks.

\begin{figure}[tb]
\centerline{\hbox{
\psfig{figure=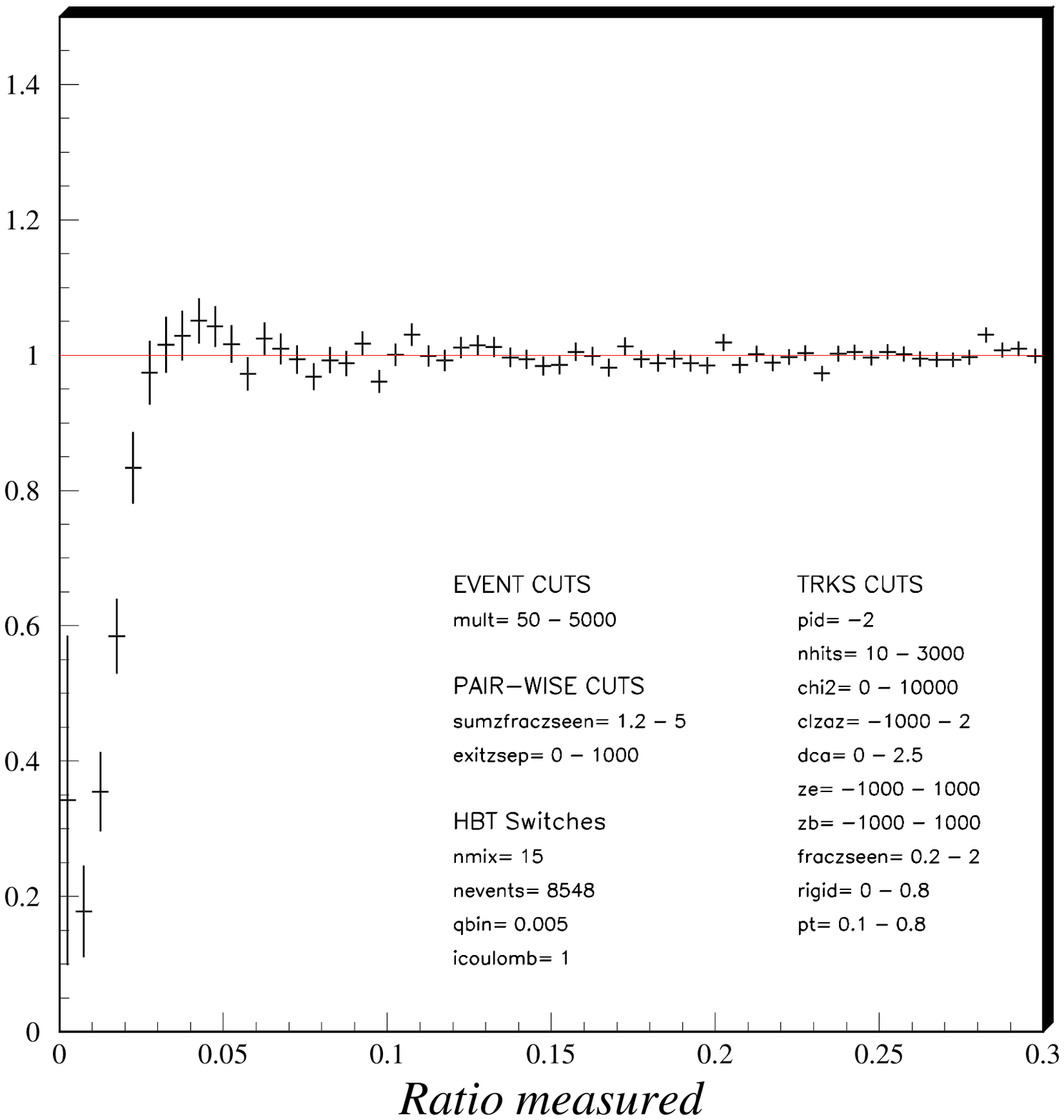,width=2.67in}
\psfig{figure=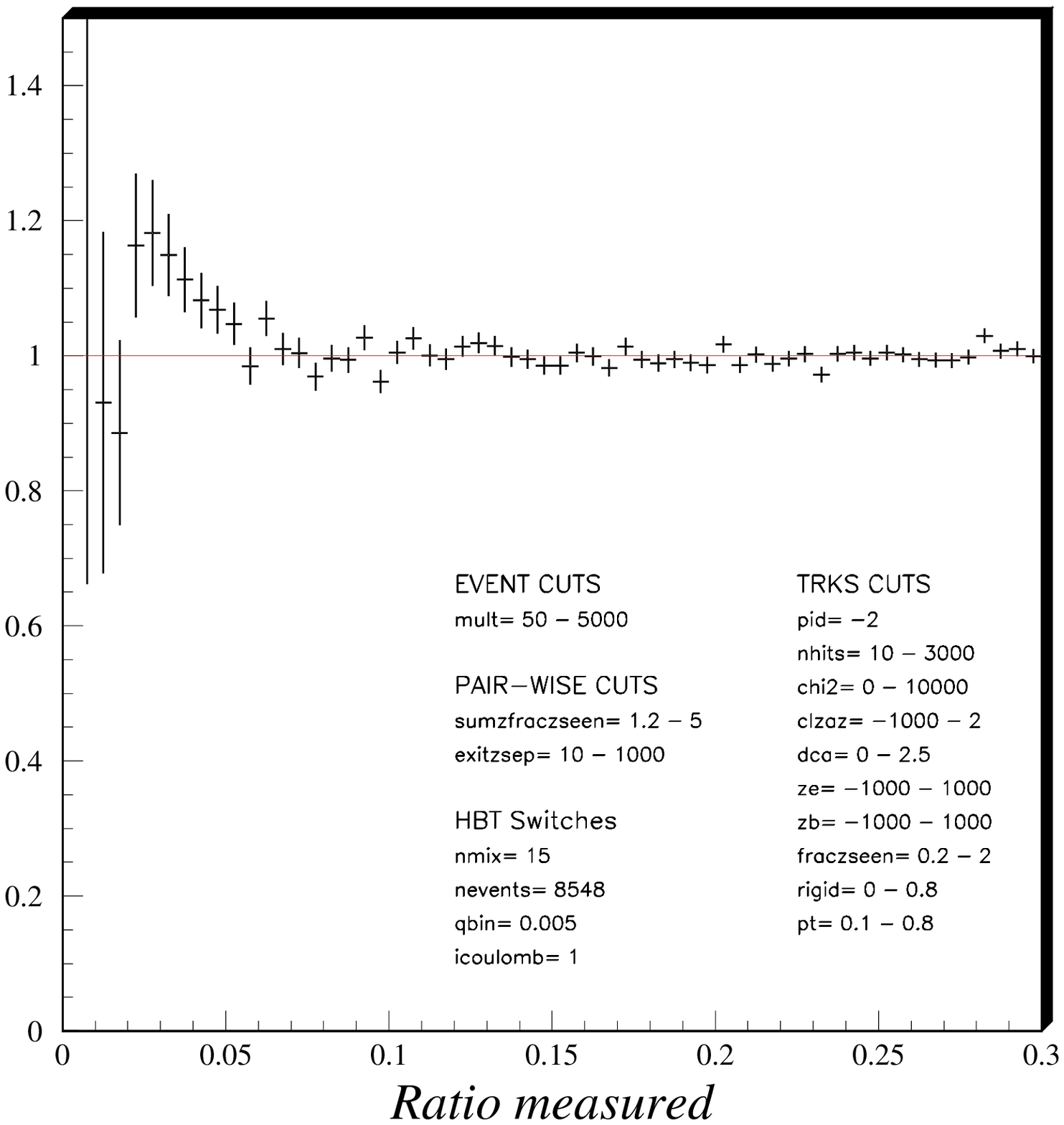,width=2.67in}
}}
\centerline{\hbox{
\psfig{figure=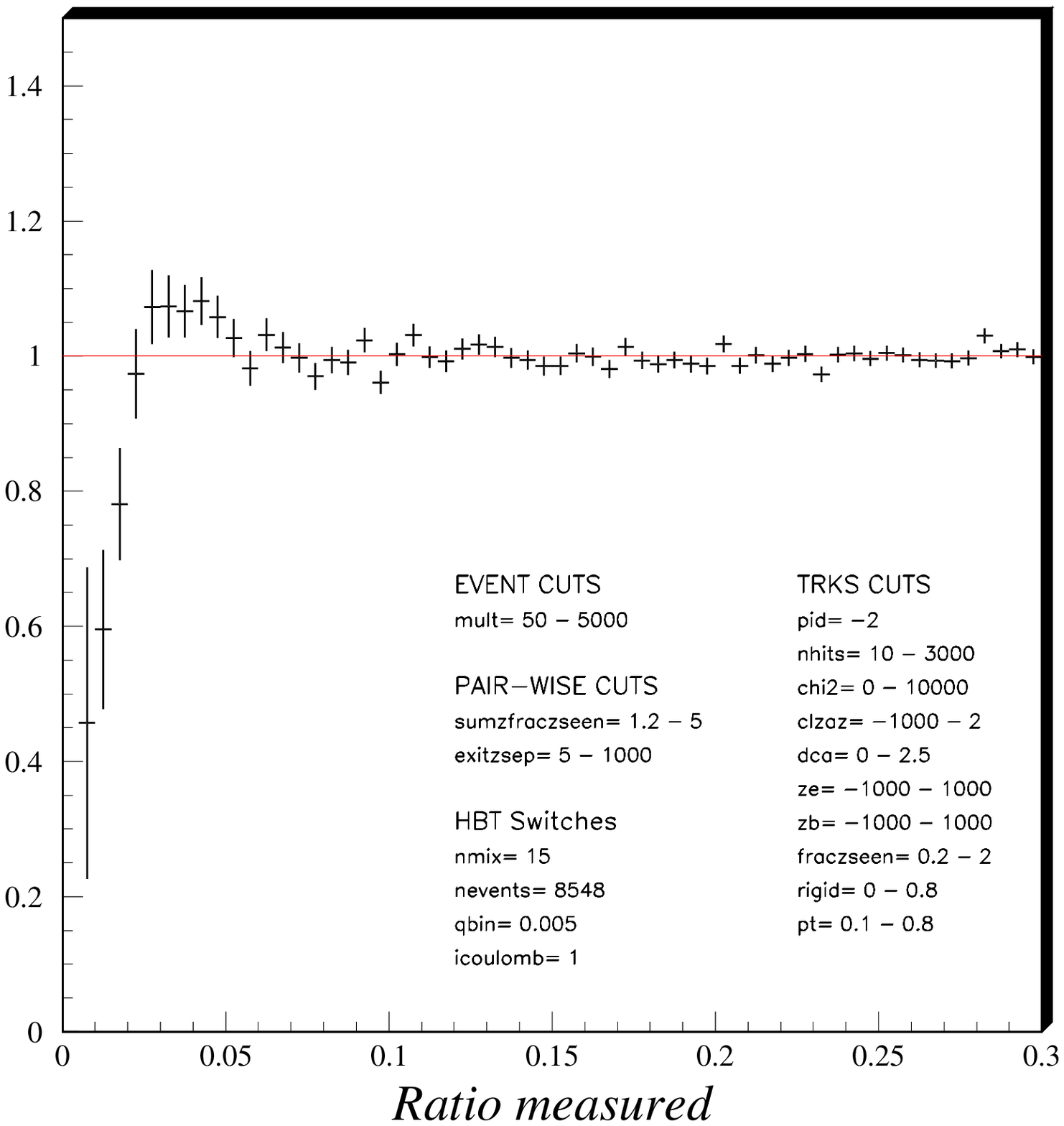,width=2.15in}
\psfig{figure=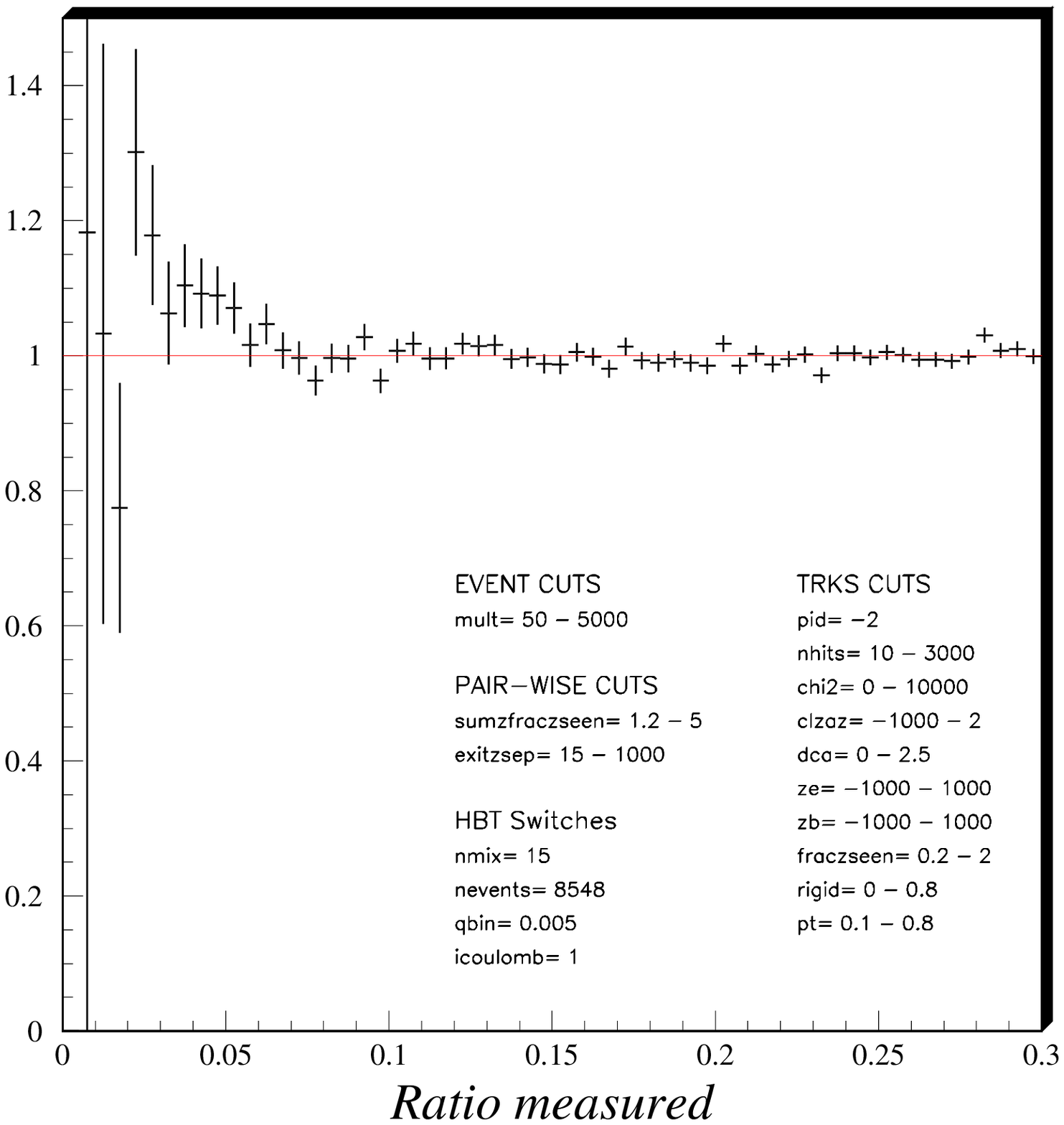,width=2.67in}
}}
\caption{After splitting is removed, the correlation function is plotted (for
the 4~AGeV data) for a cut on exit separation of 0 cm (top left), 5 cm (bottom left)
10 cm (top right) and 15 cm (bottom right).}
\label{fig:exitsep0-15}
\end{figure}

The solution implemented is to require that the {\it sum} of the reconstructed
fraction of the pair is greater than 100\%.
This removes split tracks, while preserving true low-$Q_{inv}$ pairs.
The correlation function with this cut is shown in the upper left panel
of Figure~\ref{fig:exitsep0-15}.

\subsection{Merged Tracks}

The large hole at low $Q_{inv}$ in the top left panel of 
Figure~\ref{fig:exitsep0-15} is due to track
merging-- the situation in the right panel of Figure~\ref{fig:nopaircuts}
taken to the extreme.
Merging affects pairs in the real distribution, but not in the background.
Since the merged tracks (affecting the real distribution $R$)
cannot be recovered, the goal is to remove pairs from the background
distribution $B$ which {\it would} merge if the pions came from the same event.

To this end, we cut on the distance between the projected points at which
the two tracks exit the TPC.  
For the E895 geometry,
cutting on the distance between exit points is superior to, say, cutting on the
distance between the tracks at some fixed plane (as is appropriate for
a different detector geometry\refnote{\cite{E877HBT}}), since the
tracks can exit through the any of the six sides of the rectangular detector.
A close pair that is separated by 2 cm at an intermediate plane would be
resolvable if the pions were to
pass through 40 more padrows later on, while it would be unresolvable if they
exited the TPC just after the cut plane.

As seen in Figure~\ref{fig:exitsep0-15},
raising the exit separation cut from 0 to 10 cm
reduces the low-$Q_{inv}$ hole, while further increases
only reduce statistics.

\section{CORRECTIONS APPLIED TO THE CORRELATION FUNCTION}

\vskip -1pc
\subsection{Coulomb Correction}
\label{sec:coulomb}

To measure the pure Bose-Einstein correlation,
it is common practice to ``correct'' for the Coulomb repulsion between the
charged pions when constructing the correlation function.  We apply this
pair-wise correction $G$ to the pairs in the background, 
so the measured correlation becomes

\begin{equation}
C(\kvec{1},\kvec{2}) = \frac{R(\kvec{1},\kvec{2})}
                {B(\kvec{1},\kvec{2})\times G(\kvec{1},\kvec{2})}
\label{eq:correl_gamow}
\end{equation}

In the present analysis, we use the standard Gamow function\refnote{\cite{GyulassyGamow}}
for $G$.
The Gamow function is known to over-estimate the Coulomb correction for
large source sizes\refnote{\cite{pratt:coulwave}}.
The more correct Coulomb correction
obtained by integrating the Coulomb wavefunction is expected
to change fit parameters by $\sim$10\%\refnote{\cite{E877HBT,pratt:coulwave}},
as compared to fits using the Gamow correction.
However,
presentation of results with the Gamow correction are useful for comparison
to other correlation analyses, which often use this correction.

\begin{figure}[tb]
\centerline{\hbox{
\psfig{figure=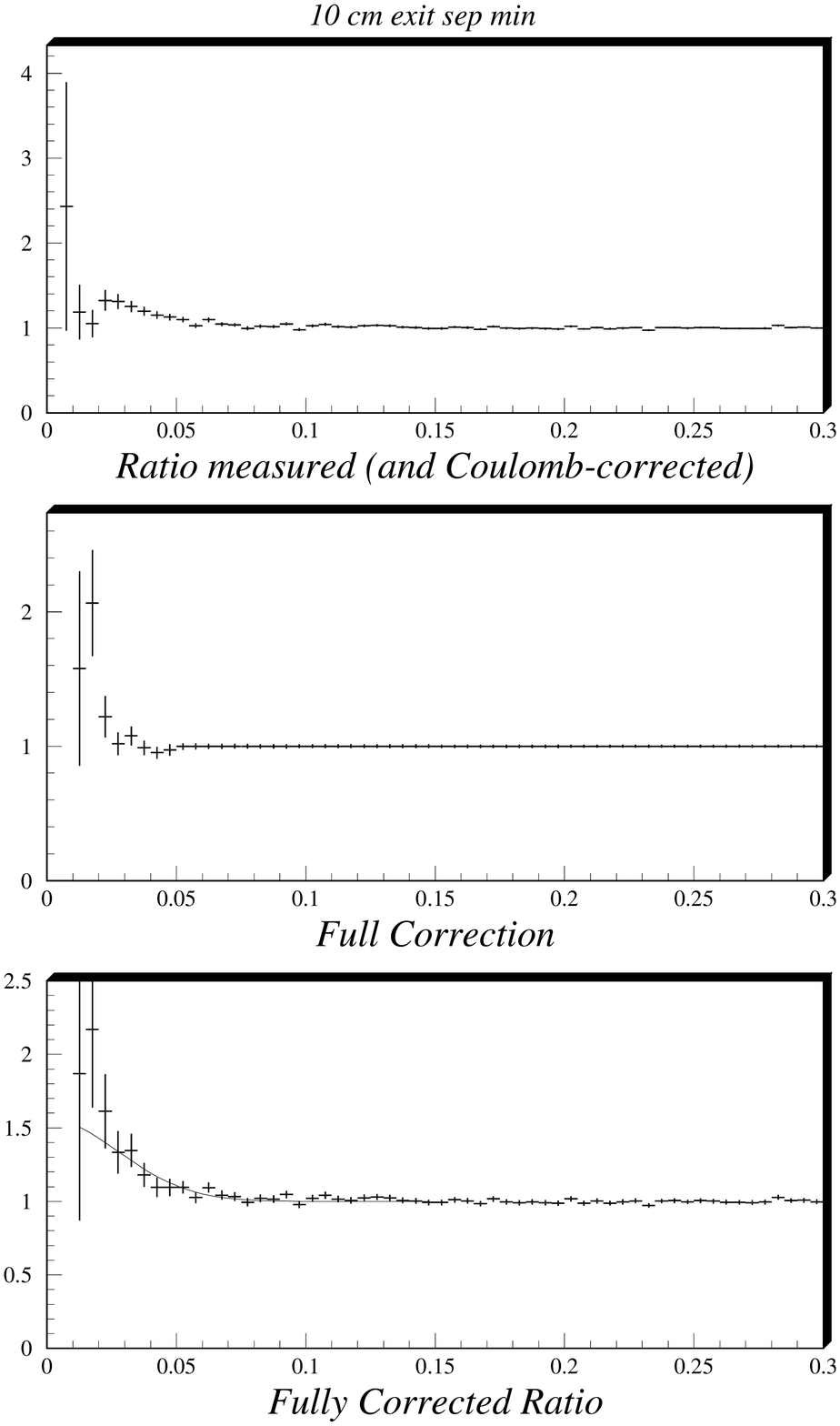,height=4.0in}
\psfig{figure=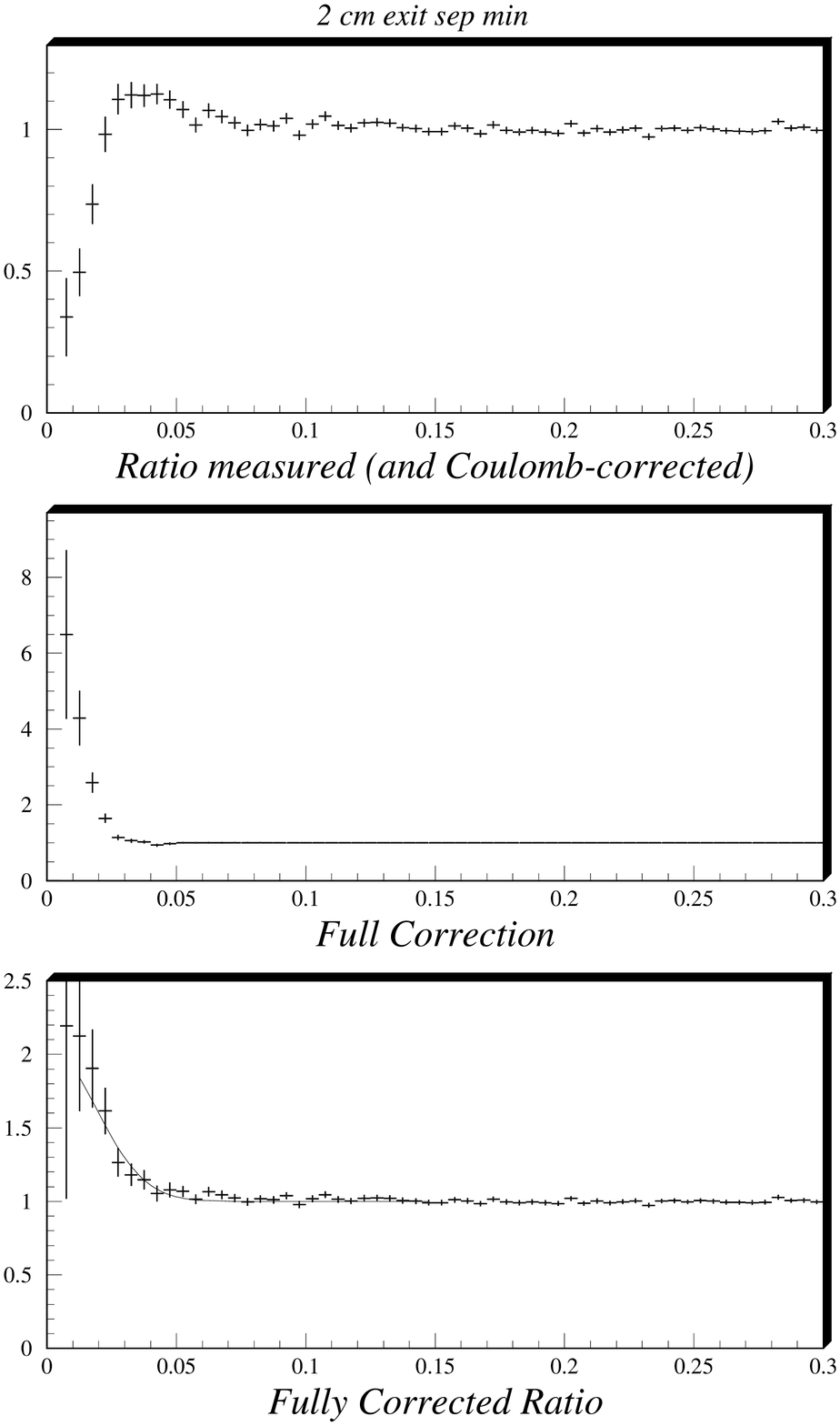,height=4.0in}
}}
\caption{Correlation functions for a 10 cm (left) and 2 cm exit
separation cut (right) from the 4~AGeV data.  Top panels show the
correlation only corrected by the Gamow function.  Middle panels show the 2-track
acceptance correction discussed in the text.  Bottom panels show the preliminary
corrected correlation functions.}
\label{fig:exitsep10and2-3panel}
\end{figure}

With all cuts and Gamow correction applied, the correlation function 
for the 4~AGeV data is shown in the
top left panel of Figure~\ref{fig:exitsep10and2-3panel}.

\subsection{Simulations and Acceptance Correction}
\newcommand{\kvp}[1]{\mbox{$\vec{k}_{#1}^{\prime}$}}

\begin{figure}[t]
\centerline{\hbox{
\psfig{figure=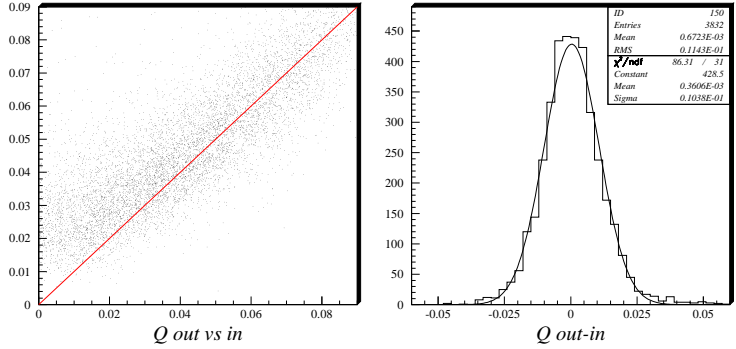,height=2.8in}
}}
\caption{Simulation results for low-$Q_{inv}$ pairs.
{\bf Left:} For very
low relative momenta, the reconstructed $Q_{inv}$ is distorted due to phase space.
\newline
{\bf Right:}Above $Q_{inv}$=40~MeV/c,
the relative momentum is resolved with a resolution of $\sim$10 MeV/c.
}
\label{fig:qres}
\end{figure}

Even with the pair-wise cuts, detailed simulations show that two-track
acceptance effects persist; these include some residual track splitting
and merging.  However the larger effect on the correlation function
is the finite resolution and distortion of the relative
momentum.  For example, for some set of cuts, Figure~\ref{fig:qres}
shows the relative momentum resolution for simulated pairs of pions.
Above $Q_{inv}\sim$40 MeV/c, the reconstructed $Q_{inv}$ tracks with
the input $Q_{inv}$, with a resolution of about 10 MeV/c.  For very
low $Q_{inv}$, the relative momentum is distorted due to phase space and
2-track efficiency effects.  These simulations include multiple scattering
in the detector entrance window, as well as in the gas.  The simulation
is done at the pixel level, and the full
pattern recognition and reconstruction code is run.

The magnitude and character of these effects depend strongly
on (1) \kvec{1}\ and \kvec{2}\ (in six dimensions!) themselves, due to the
irregular geometry of the TPC, and (2) the particular track-wise and pair-wise
cuts used in a given analysis.  Rather than attempting to parametrize these
effects as a function of this huge number of variables, we employ detailed
simulations directly in our HBT analysis.  The technique described here follows
closely that of the NA44 collaboration\refnote{\cite{NA44-accept}}.

The singles momentum distribution (for some impact parameter cut)
is fit with a thermal source distribution.  This distribution is then sampled
to give pion pairs at low-$Q_{inv}$, which are then embedded at the pixel level
into real data events.  This method ensures that the true noise distribution
and track density present in the data affects the reconstruction of simulated
particles in the same way as it does the measured particles.
It also ensures that the same (\kvec{1},\kvec{2}) distribution is used in the
corrections as in the data analysis.
For each pair input to the simulation, there are two input momenta
(\kvec{1},\kvec{2}), and $n$ output (reconstructed) momenta
(\kvp{1},\kvp{2},\ldots,\kvp{n}).  An ideal detector and event reconstruction
would yield $n=2$ and \kvp{i}=\kvec{i}.  The purpose of the correction is
to remove effects of the deviation from this idealization.
It should be noted
that effects of track merging and splitting ($n\neq 2$) and momentum resolution
($\kvp{i}\neq\kvec{i}$) will depend on the cuts we use.  Thus, we would require
our correction to track with any changes in cut values.

The acceptance correction is then defined as

\begin{equation}
K = \frac{C(ideal)}{C(reconstructed)}
  = \frac{\frac{R(\kvec{1},\kvec{2})}{B(\kvec{1},\kvec{2})}}
         {\frac{R(\kvp{1},\kvp{2})}{B(\kvp{1},\kvp{2})}}
\label{eq:acceptcor}
\end{equation}

\noindent
where $R(\kvec{1},\kvec{2})$ is the real distribution of simulated (input)
pairs weighted by the correlation function:

\begin{equation}
R(\kvec{1},\kvec{2}) = \frac{d^6N}{d^3\kvec{1}d^3\kvec{2}}\times C(\kvec{1},\kvec{2})
\label{eq:input_num}
\end{equation}

\noindent
and $B(\kvec{1},\kvec{2})$ is the background distribution

\begin{equation}
B(\kvec{1},\kvec{2}) = \frac{d^6N}{d^3\kvec{1}d^3\kvec{2}}
\label{eq:input_den}
\end{equation}

\noindent
Note that use of these relations to construct the numerator
in equation \ref{eq:acceptcor} (as opposed to simply using $C(\kvec{1},\kvec{2})$),
allows an accounting for the statistical error.
In forming the ideal correlation function, the only cuts applied to the
input tracks are phase-space cuts (rapidity and $p_T$)-- no track quality
cuts.

The denominator of equation~\ref{eq:acceptcor} is
formed with the reconstructed particles,
with momenta \kvp{i}.  
As mentioned above, the number $n$ of these particles for
one input pair, will depend on the particular value of
track-wise and pair-wise cuts used.
In forming this reconstructed correlation function,
the same cuts are applied as to the measured data.

 $R(\kvp{1},\kvp{2})$ is formed by binning the distribution according to the
reconstructed momenta \kvp{i}, but {\it weighted} by the {\it true} Bose-Einstein
correlation function, which is a function of the input (``true'') momenta \kvec{i}.

\begin{equation}
R(\kvp{1},\kvp{2}) = \frac{d^6N}{d^3\kvp{1}d^3\kvp{2}}\times C(\kvec{1},\kvec{2})
\label{eq:output_num}
\end{equation}

\noindent
In the simulation, then, we must keep track of which input particle (\kvec{i})
gives rise to a reconstructed track (\kvp{i}).

\subsection{Correction to the Coulomb Correction}

To complete the acceptance correction (Equation~\ref{eq:acceptcor}), we must
calculate the background distribution of reconstructed particles.  In doing
so, however, we account for the fact that the Coulomb correction $G$
applied to the measured background distribution (Equation~\ref{eq:correl_gamow})
is calculated using the measured $Q_{inv}$, and not the
true $Q_{inv}$\refnote{\cite{NA44-accept,E877HBT}}.
We account for this finite resolution effect by constructing the reconstructed
background as

\begin{equation}
B(\kvp{1},\kvp{2}) = \frac{d^6N}{d^3\kvp{1}d^3\kvp{2}}\times
                        \frac{G(\kvec{1},\kvec{2})}{G(\kvp{1},\kvp{2})}
\label{eq:output_den}
\end{equation}

\noindent
where $G$ is the Coulomb correction (here, just the Gamow function).

Note that these considerations make it clear that the corrections for Coulomb
repulsion and detector acceptance cannot be factorized.

\subsection{Notes}

We see that in constructing the acceptance correction, the correlation function
was used as a weight.  But this is the very thing we are trying to measure!
Again following the formalism of the NA44 collaboration\refnote{\cite{NA44-accept}},
we employ an iterative approach to the problem.

We assume a Gaussian source distribution $\rho(r)\sim e^{-r^2/R^2}$, which would
lead to a Gaussian correlation function

\begin{equation}
C(Q_{inv})=1+\lambda e^{-(Q_{inv}\cdot R)^2}
\label{eq:parametrization}
\end{equation}

\noindent
where $\lambda$ is the so-called coherence parameter.  At the first iteration,
we assume some reasonable values for $R$ and $\lambda$, and construct the correction.
We then fit the corrected correlation function with the form (\ref{eq:parametrization}).
This gives new values of $R$ and $\lambda$, which are then used as inputs for
the next iteration.  It is found that the fit parameters stabilize within about
4 iterations, and are robust against variations in the initial guess.

This proceedure will become even more important when we use the full Coulomb wave
integration (instead of Gamow), since then the Coulomb correction itself depends
on the source distribution.

Finally, it may be possible to do away with the assumption of a Gaussian source
altogether, by taking the Fourier transform of the correlation function to
extract the source distribution\refnote{\cite{Brown_Pawel}} in the iterative
process.  We plan to try this as a next step.

\section{RESULTS}

In Figure~\ref{fig:exitsep10and2-3panel} are shown
the raw (but Gamow-corrected) correlation function, the acceptance correction,
and the corrected correlation
function for two different cuts on the exit separation for the 4~AGeV data.
As noted above, the low-$Q_{inv}$ hole in the raw correlation function, due
to track merging, varies with the value of this cut.  However, when the same cuts
are applied to the reconstructed tracks from the simulation, the acceptance
correction is seen to change as well-- a larger correction is calculated, as expected,
for the looser cut on exit separation.  Finally, when the correction is applied, it
is seen that the correlation functions for the two different cuts agree.

\begin{figure}[tb]
\centerline{\hbox{
\psfig{figure=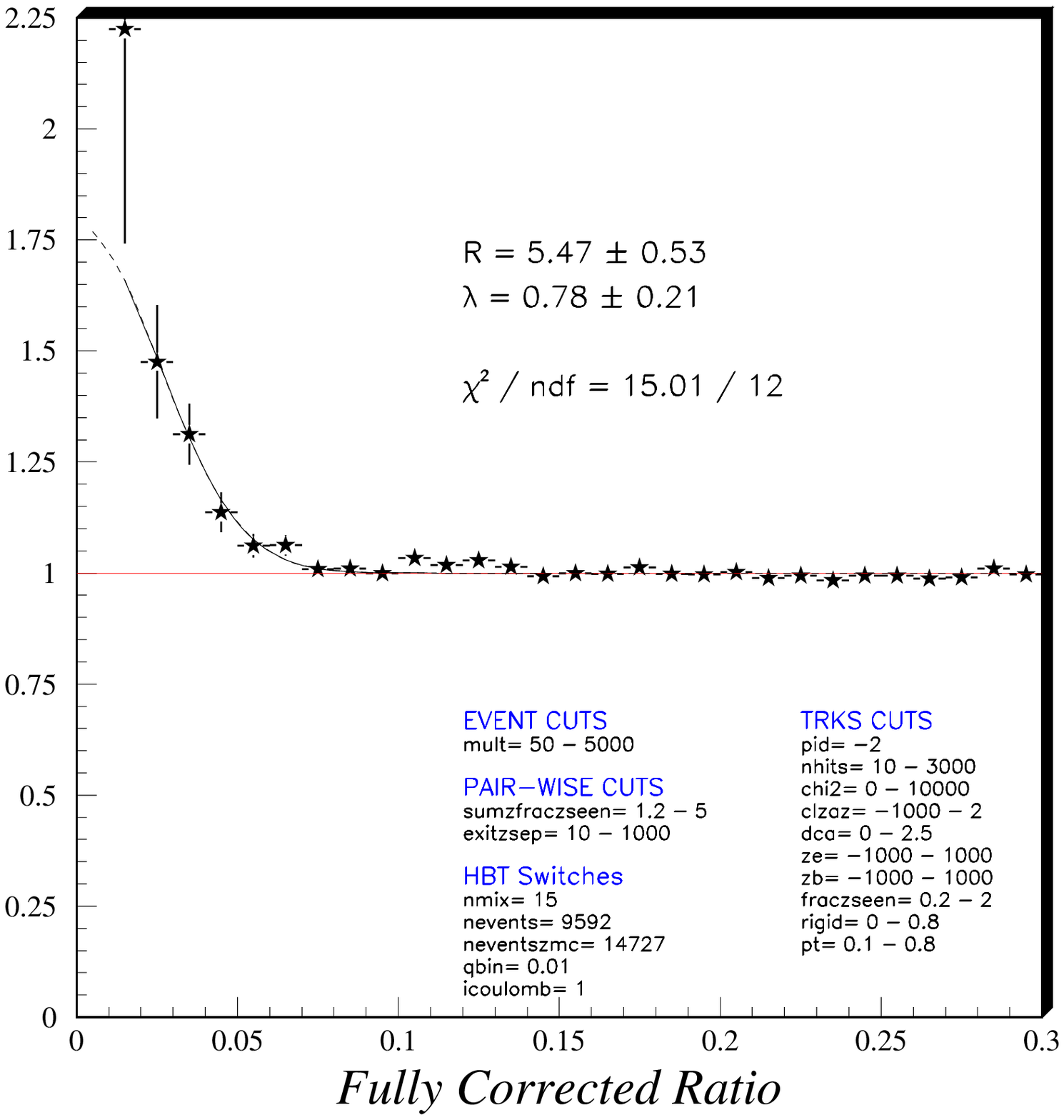,height=3.1in}
\psfig{figure=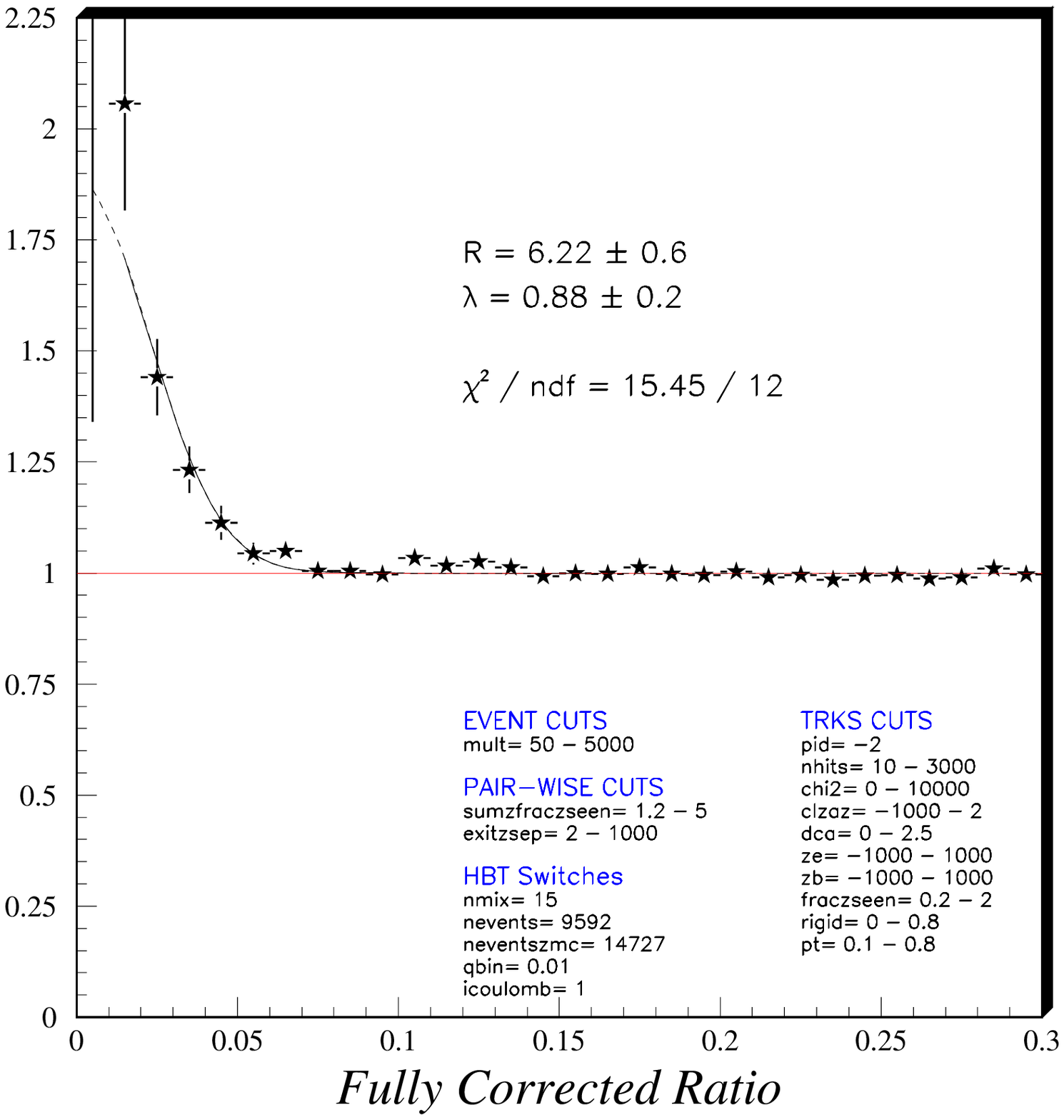,height=3.1in}
}}
\caption{Preliminary fully corrected correlation functions for the 4 A GeV data are shown,
for an exit separation cut of 10 cm (left) and 2 cm (right).}
\label{fig:4gevexit2-4gevexit10}
\end{figure}

Variations in other cut parameters also produce different raw correlation functions,
and different correction factors.  However, the corrected correlation function is
robust against reasonable variations.  This stability
gives us confidence in our ability to remove the nontrivial effects of the detector.
The fully corrected correlation functions for the 4 A GeV data are shown in
Figure~\ref{fig:4gevexit2-4gevexit10}.
Gaussian fits to the correlation function with the form
of Equation~\ref{eq:parametrization} give an invariant radius of
 $R = 5.8 \pm 0.7$ fm and incoherence parameter $\lambda = 0.83 \pm 0.25$.

\begin{figure}[tb]
\centerline{\hbox{
\psfig{figure=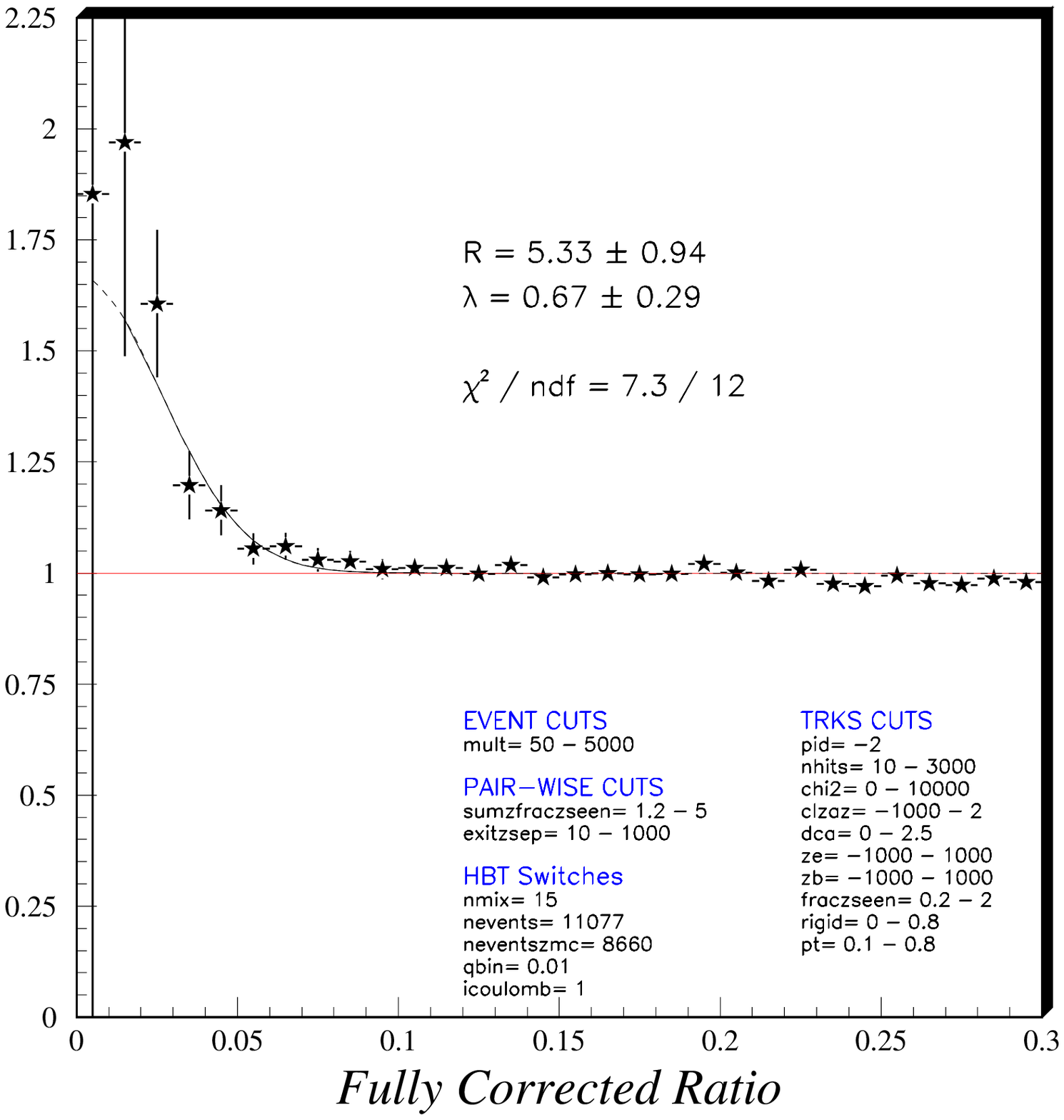,height=3.1in}
\psfig{figure=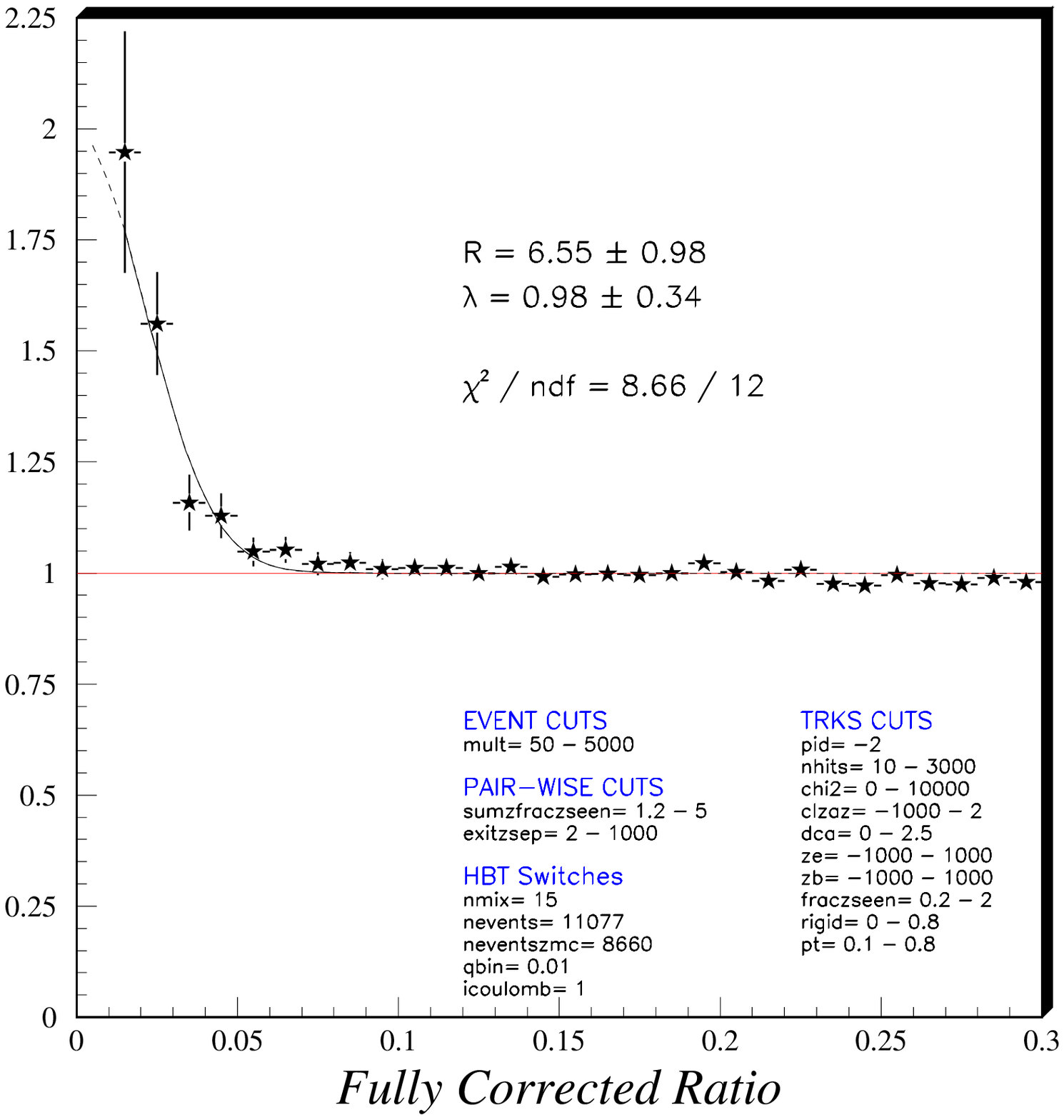,height=3.1in}
}}
\caption{Preliminary fully corrected correlation functions for the 2 A GeV data are shown,
for an exit separation cut of 10 cm (left) and 2 cm (right).}
\label{fig:2gevexit2-2gevexit10}
\end{figure}

The same stability is seen in the analysis of the 2~AGeV data.
Figure~\ref{fig:2gevexit2-2gevexit10} shows the corrected
correlation function for two different values of the exit separation cut.
Although the data points themselves fall almost on top of each other, the Gaussian
fit parameters are seen to be very sensitive to small variations.  These
variations in the fit parameters can be treated as a systematic error.
Fits, shown in the figure, give  $R = 5.9 \pm 1.2$ fm and $\lambda = 0.83 \pm 0.34$.

The radius parameters extracted (often called $R_{inv}$) are consistent with
those extracted at 10 AGeV\refnote{\cite{E877HBT,E802HBT}}, when the Gamow
Coulomb correction is used.
Meanwhile the $\lambda$ parameters presented here higher than the values
(0.45-0.6) obtained at the higher energy.  This may be expected due to the
decreased role of long-lived resonances, which tend to
reduce $\lambda$\refnote{\cite{Csorgo-lambda}}.  However, with
the statistics used in this analysis, error bars are too large to confirm
a difference.

Not shown here are the results for the 8~AGeV data.  For this set, the corrected
correlation function was {\it not} stable against variations in the cuts.
We are currently tracking down the source of this problem, which appears to
be a tracking error when the track density gets very high.

\section{SUMMARY}

Details of the E895 $\pi^-$ correlation analysis have been presented.  Along with
track-wise criteria to select well constructed pions, pair-wise cuts are neccessary
to reduce the two-track detector acceptance effects.  Remaining two-track effects
can be corrected through use of detailed simulations, in which simulated low-$Q_{inv}$
pion pairs are embedded in real data at the pixel level, and then reconstructed.
The uncorrected correlations depend strongly on the cut values used, as do the
corrections obtained with the simulation.  However, the corrected correlation functions
are largely robust against variations in the cuts.

Preliminary one-dimensional correlation functions have been presented for Au+Au
collisions at 2 and 4 AGeV, with a medium-bias impact parameter distribution.  Gaussian
fits to these functions yield radius parameters consistent with those obtained at the
maximum AGS energy, while the intercept parameter appears larger, perhaps a sign
that long-lived resonances play less of a role at these energies.

A problem currently under study is that the stability of the correlations
at 2 and 4 AGeV is not seen in our analysis at 8 AGeV.
Resolution of this problem, greater statistics at all energies, and
a multidimensional analysis will
be neccessary to definitively say that HBT parameters show no sharp behavior as
the beam energy is changed.


\section{ACKNOWLEGEMENTS}

The author thanks Drs. David Hardtke and Thomas Humanic for useful
discussions on the acceptance correction.

This work supported by the Director, Office of Energy Research, Office of Basic
Energy Sciences, Nuclear Science Division, of the U.S. Department of Energy under
contrack DE-AC03-76SF00098 and grants DE-FG02-89ER40531, DE-FG02-88ER40408, DE-FG02-87ER40324,
by the U.S. National Science Foundation under grants PHY-9722653, PHY-9601271,
PHY-9225096, and by the University of Auckland Research Committee, NZ/USA Cooperative
Science Programme CSP 95/33.

\end{document}